\documentclass[preprint,showpacs,12pt,nofootinbib]{revtex4}

\usepackage{amsmath,amssymb,amsfonts}
\usepackage{graphicx}% Include figure files
\usepackage{hhline}
\usepackage{bm}
\usepackage{amsmath}
\usepackage{epsfig}
\usepackage{graphicx}

\usepackage{tabularx}
\usepackage{lscape}
\usepackage{rotating}
\usepackage{pstricks}

\usepackage{footnote}
\usepackage{hyperref}

\usepackage{ulem}

\begin{document}
\title{Phase space of electron- and muon-neutrino and antineutrino scattering off nuclei}

\author{M. Martini}
\affiliation{IPSA-DRII,  63 boulevard de Brandebourg, 94200 Ivry-sur-Seine, France}
\affiliation{Sorbonne Universit\'e, CNRS/IN2P3, Laboratoire
de Physique Nucl\'eaire et de Hautes Energies (LPNHE), Paris, France}
\author{M. Ericson} 
\affiliation{Univ Lyon, Univ Claude Bernard Lyon 1, CNRS/IN2P3, IP2I Lyon, UMR 5822, F-69622, Villeurbanne, France}
\affiliation{Theory division, CERN, CH-12111 Geneva }
\author{G. Chanfray} 
\affiliation{Univ Lyon, Univ Claude Bernard Lyon 1, CNRS/IN2P3, IP2I Lyon, UMR 5822, F-69622, Villeurbanne, France}

\begin{abstract}

We discuss the electron and muon neutrino and antineutrino double differential cross sections on carbon in the quasielastic as well as in the multinucleon and one pion production channels. By projecting them in the transferred  momentum - transferred energy plane and in the neutrino energy - lepton scattering angle plane, as well as by performing simple considerations on the position of the quasielastic and Delta peaks and on their broadening, we explain the surprising dominance of the muon neutrino and antineutrino cross sections over the electron ones in particular kinematical conditions.

\end{abstract}

\maketitle
\section{Introduction}
One of the main objectives of present \cite{T2K:2019bcf, NOvA:2019cyt} and future \cite{Hyper-Kamiokande:2018ofw,DUNE:2015lol,Alekou:2022emd} accelerator-based neutrino oscillation experiments is the search for the charge-parity (CP) violation in the leptonic sector. 
The best way to observe this phenomenon would be the measurement of a different appearance probability for electron neutrinos and electron antineutrinos from intense beams of muon neutrinos and muon antineutrinos. 
The next generation long-baseline (LBL) experiments will have unprecedented statistics of detected neutrinos thanks to intense beams and huge detector size. However these features are not sufficient to guarantee their success in the potential discovery of CP violation. 
%LBL neutrino experiments exploit neutrino-nucleus interactions to measure neutrino oscillations. 
In contrast with old bubble-chamber experiments, where the interaction of the neutrinos occurs with hydrogen, 
the use of relatively heavy nuclear targets (carbon, oxygen, argon), while allowing for a substantial increase of the event rate,
%, entails non trivial problems, since the interpretation of the data 
requires a quantitative description of the nuclear response to weak interactions \cite{Katori:2016yel,NuSTEC:2017hzk}. 
%In view of the expected experimental accuracy, 
A precise and simultaneous knowledge of the  $\nu_\mu$, $\nu_e$, $\bar{\nu}_\mu$ and $\bar{\nu}_e$  cross sections on the target nucleus will be indeed crucial for the success of the LBL experiments. 

In this connection, the last fifteen years have been characterized by numerous $\nu_\mu$ and $\bar{\nu}_\mu$ cross sections measurements. On the contrary the equivalent data for $\nu_e$, and $\bar{\nu}_e$ are scarce and unlikely to reach the same level of precision as the $\nu_\mu$ and $\bar{\nu}_\mu$ ones. A theoretical investigation on the difference between electron and muon cross sections is hence particularly important. 

In charged current processes
\begin{equation}
    \nu_l + A \to l^{-}+X
\end{equation}
\begin{equation}
    \bar{\nu}_l + A \to l^{+}+X,
\end{equation}
where $l$ denotes the generic flavour (which can be $e$ or $\mu$), 
$\nu_e$ ($\bar{\nu}_e$) cross sections are expected to be larger than the $\nu_\mu$ ($\bar{\nu}_\mu$) ones due to the differences in the mass of the outgoing charged lepton, which imply different kinematic limits. This is certainly true for the total neutrino cross section $\sigma_{\nu_l}$ as a function of the neutrino energy $E_{\nu_l}$. However this hierarchy can be opposite in specific kinematical conditions in the case of differential cross sections, as $\frac{d \sigma}{d\cos\theta}$, where $\theta$ is the lepton scattering angle, and $\frac{d^2 \sigma}{d E_l d\cos\theta}$, where $E_l$ is the charged-lepton energy, or equivalently $\frac{d^2 \sigma}{d\omega d\cos\theta}$, where $\omega$ is the transferred energy, $\omega=E_{\nu_l}-E_l$. 

This surprising inversion of the  $\nu_e$ and  $\nu_\mu$ cross section hierachy was pointed out at first in Ref. \cite{Martini:2016eec} where it was shown that for forward scattering angles the muon neutrino quasielastic differential cross sections can be larger than the corresponding electron ones, especially for low neutrino energies. This unexpected feature, and its potential important impact for the LBL neutrino oscillation results, pushed the community to perform further investigation in this direction. Hence several papers on this subject have been published \cite{Ankowski:2017yvm,Nikolakopoulos:2019qcr,Gonzalez-Jimenez:2019qhq,Dieminger:2023oin}.  
Reference \cite{Martini:2016eec} already stressed that the surprising dominance of $\nu_\mu$ over $\nu_e$ quasielastic differential cross sections at fixed kinematics for small scattering angles is related to the differences in the momentum transfer ${\bf{q}}={\bf{k}}_{\nu_l}-{\bf{k}}_{l}$ between the $\nu_e$ and $\nu_\mu$ scattering. 
%and to non-trivial dependence of momentum transfer on lepton mass. 
Reference \cite{Ankowski:2017yvm} analyzed the $(q,\omega)$ phase space available for the charged current quasielastic (CCQE) interaction and pointed out that the $\nu_\mu$ over $\nu_e$ dominance, appearing in the Fermi-Gas based and Hartree-Fock based approaches \cite{Martini:2016eec}, could no more appear by considering a spectral function approach. However in the calculations of  Ref. \cite{Ankowski:2017yvm} nucleon final-state interactions were not taken into account. References \cite{Nikolakopoulos:2019qcr,Gonzalez-Jimenez:2019qhq} used several independent mean-field based models to conclude that a proper quantum-mechanical treatment of Pauli blocking and of the final nucleon’s wave function confirms the dominance of $\nu_\mu$ over $\nu_e$  cross sections at forward lepton scattering angle. In Ref. \cite{Dieminger:2023oin} the potential for mis-modeling of $\nu_e/\nu_\mu$ and $\bar{\nu}_e/\bar{\nu}_\mu$ CCQE cross-section ratios was quantified in order to investigate its impact on neutrino oscillation experiments. In this analysis large differences between the Hartree-Fock based 
and spectral function approaches appeared in the forward scattered region and, even if less pronounced that in the Hartree-Fock case, a region where the $\nu_e/\nu_\mu<1$ appeared also in the spectral function case for small neutrino energy. Furthermore it was also shown that for the antineutrino case a region appears in the $(\theta, E_\nu)$ phase space where $\bar{\nu}_e/\bar{\nu}_\mu<1$. This happens at backward scattering angles for different theoretical models. 
%\sout{In the present work we want to complement the analyses of the mentioned previous papers \cite{Martini:2016eec,Ankowski:2017yvm,Nikolakopoulos:2019qcr,Gonzalez-Jimenez:2019qhq,Dieminger:2023oin} by discussing simple arguments and by introducing an effective way of presenting results which could be useful for future investigations.}

In the present work we want to complement the previous investigations by performing phase space and kinematical analysis 
which allows us to formulate simple and original explanations, not explicitly done in any of the previous published papers \cite{Martini:2016eec,Ankowski:2017yvm,Nikolakopoulos:2019qcr,Gonzalez-Jimenez:2019qhq,Dieminger:2023oin}, of the surprising $\nu_\mu$ over $\nu_e$ dominance in the quasielastic channel. The same analysis is generalized also to the other channels included in our approach \cite{Martini:2009uj}, \textit{i.e.} the multinucleon emission, the incoherent and coherent one pion production, marginally discussed only in Ref.\cite{Martini:2016eec} and omitted in Refs. \cite{Ankowski:2017yvm,Nikolakopoulos:2019qcr,Gonzalez-Jimenez:2019qhq,Dieminger:2023oin} which focused on quasielastic only.  

Although neutrino beams are not monochromatic, we decide to consider, as in all the previous theoretical papers on this subject \cite{Martini:2016eec, Ankowski:2017yvm,Nikolakopoulos:2019qcr,Gonzalez-Jimenez:2019qhq}, only the case of fixed neutrino energy. This variable $E_{\nu_l}$, as well as the transferred energy $\omega$ and momentum $q$, is not directly measurable and up to now the neutrino scattering community rightly privileged the flux integrated cross sections as a function of measured variables such as the lepton energy $E_l$ and the scattering angle $\theta$. However there are several reasons to consider the cross sections also in terms of fixed neutrino energy. 
First we want to confine ourselves to the same conditions as those of the previous published analyses on the same subject \cite{Martini:2016eec,Ankowski:2017yvm,Nikolakopoulos:2019qcr,Gonzalez-Jimenez:2019qhq}. Second, present and future neutrino detectors will allow more and more exclusive measurements and to know better and better the vertex activity. This, combined with more and more accurate neutrino interaction modeling, will allow to reconstruct and constrain unmeasured variables. First examples of experimental cross sections shown as a function of directly-unmeasurable variables, such as $\sigma(E_{\nu_\mu})$, $\frac{d \sigma}{d \omega}$ \cite{MicroBooNE:2021sfa} and, more recently, $\frac{d^2 \sigma(E_{\nu_\mu})}{d k_\mu d \cos \theta}$ \cite{MicroBooNE:2023foc} already appeared. Finally it allows theoretical analyses and effective visualization of different cross sections behaviour, as we show in the following.

\section{Quasielastic channel}
Let us start by representing in the $(q,\omega)$ plane the charged current quasielastic double differential cross sections $\frac{d^2 \sigma (E_{\nu_l})}{d\omega d\cos\theta}$ on carbon for all the values of the scattering angle and for fixed values of the neutrino energy.

\begin{figure}
\begin{center}
  \includegraphics[width=16cm]{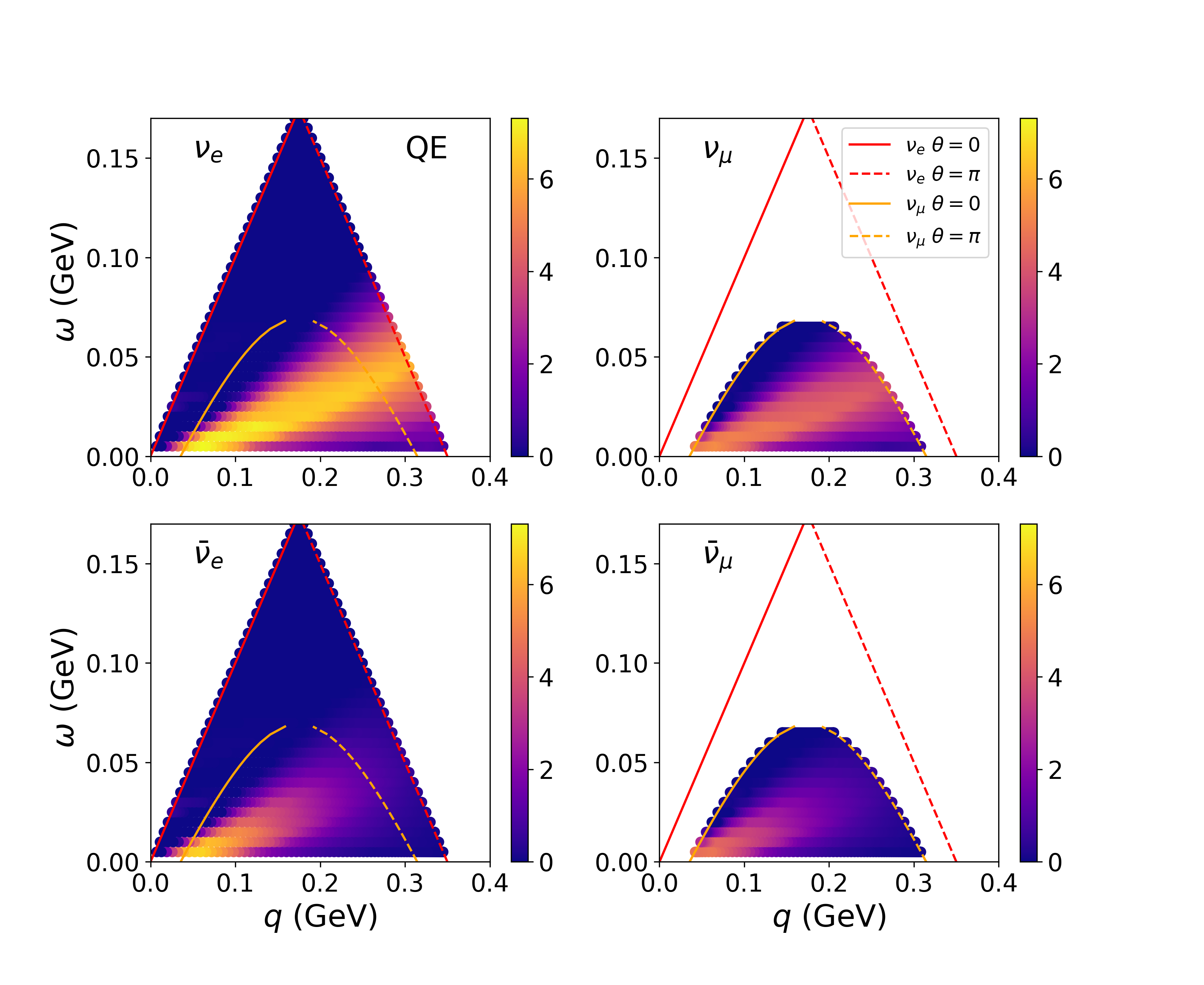}
\caption{Projection in the $(q,\omega)$ plane of the $\nu_e$, $\nu_\mu$, $\bar{\nu}_e$ and $\bar{\nu}_\mu$ charged current 
quasielastic double differential cross
section on carbon $\frac{d^2 \sigma (E_{\nu_l})}{d\omega d\cos\theta}$  for the fixed value of the neutrino energy $E_{\nu_l}=175$ MeV and 
for all the values of the scattering angle.  The strength of the cross section, in units of $10^{-38}$ cm$^2$/GeV, is given by the colour scale. The curves corresponding to the $\omega-q$ relation given by Eq. (\ref{eq_q_relation}) for fixed values of the neutrino energy and charged lepton mass for the two extreme values of the lepton scattering angle, $\theta=0$ and $\theta=\pi$, are also plotted.}
\label{fig_Enu175_mu_e_nu_anti}
\end{center}
\end{figure}

\begin{figure}
\begin{center}
  \includegraphics[width=16cm]{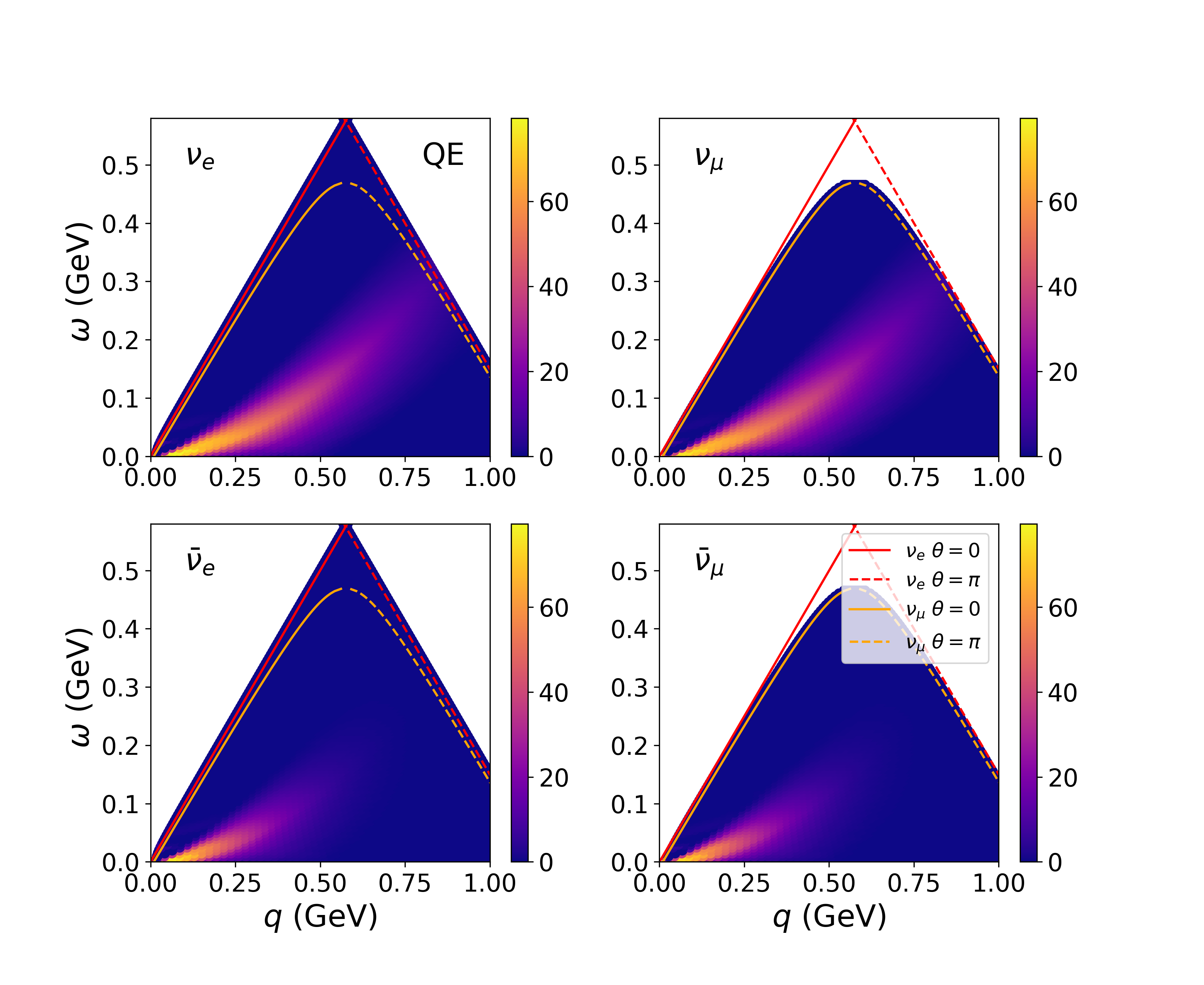}
\caption{The same as Fig.\ref{fig_Enu175_mu_e_nu_anti} but for $E_{\nu_l}=575$ MeV.}
\label{fig_Enu575_mu_e_nu_anti}
\end{center}
\end{figure}

We remind that the values of $q=|{\bf{q}}|$ are obtained by the formula
\begin{equation}
\label{eq_q_relation}
    q = \sqrt{E^2_{\nu_l}+k_l^2-2E_{\nu_l}k_l\cos\theta},
\end{equation}
where
\begin{equation}
\label{eq_kl}
   k_l^2= E^2_{l}-m^2_l=(E_{\nu_l}-\omega)^2-m^2_l.
\end{equation}
Once $E_{\nu_l},~\omega$ and $\cos\theta$ are fixed, $q$ is  determined, hence it is possible to project $\frac{d^2 \sigma}{d\omega d\cos\theta}$ in the $(q,\omega)$ plane, the strength of the cross section being represented by a colour chart. These cross sections, referring exclusively to the genuine CCQE channel, are shown in Figs.\ref{fig_Enu175_mu_e_nu_anti} and \ref{fig_Enu575_mu_e_nu_anti} for the four cases $\nu_e$, $\nu_\mu$, $\bar{\nu}_e$ and $\bar{\nu}_\mu$, for the neutrino energies $E_{\nu_l}=175$ MeV and $E_{\nu_l}=575$ MeV, respectively. In the figures we show also the curves corresponding to the $\omega-q$ relation given by Eq.(\ref{eq_q_relation}) for fixed values of the neutrino energy and charged lepton mass for the two extreme values of the lepton scattering angle $\theta=0$ and $\theta=\pi$. These curves delimit the available phase space.  

\begin{figure}
\begin{center}
  \includegraphics[width=8cm]{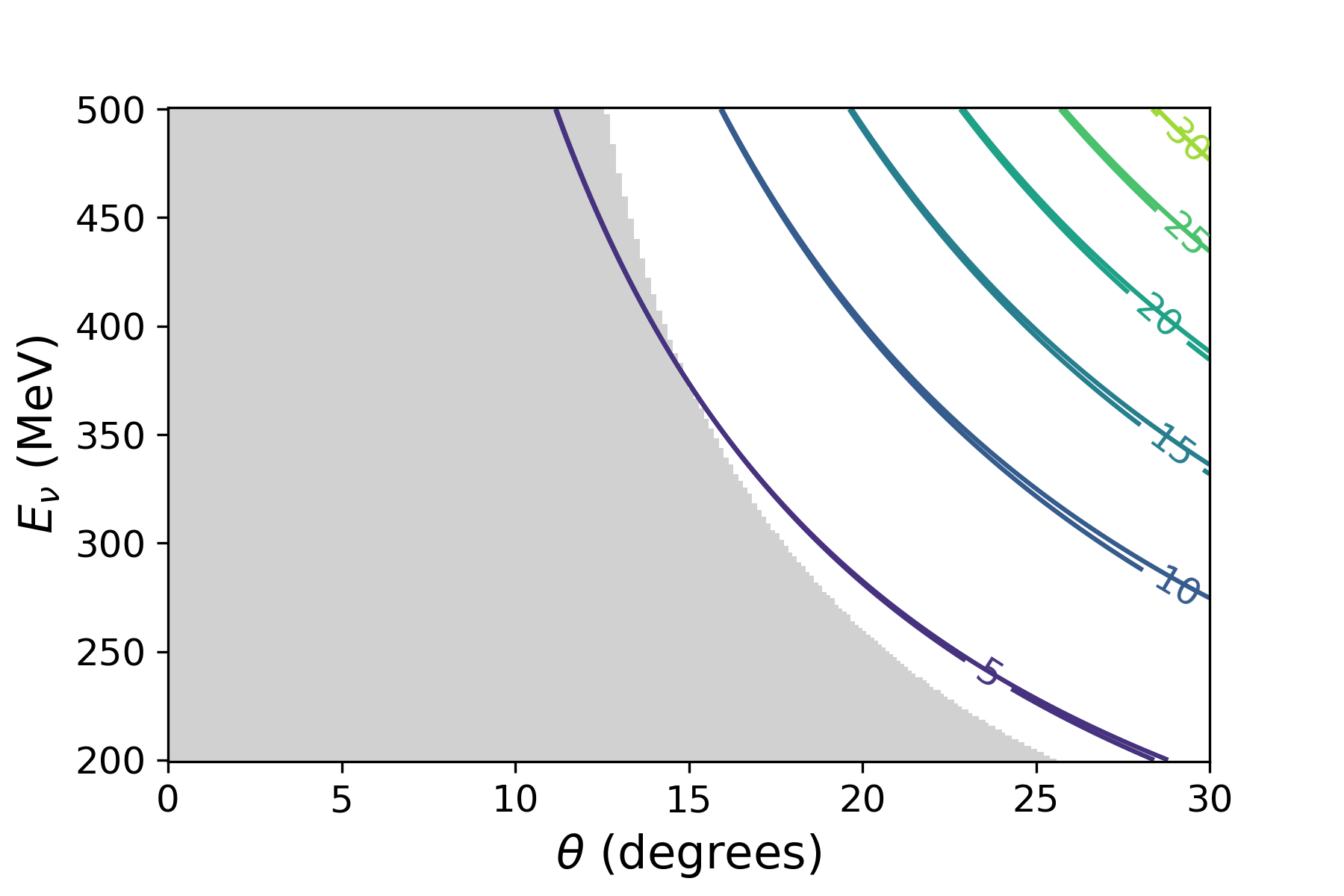}
  \includegraphics[width=8cm]{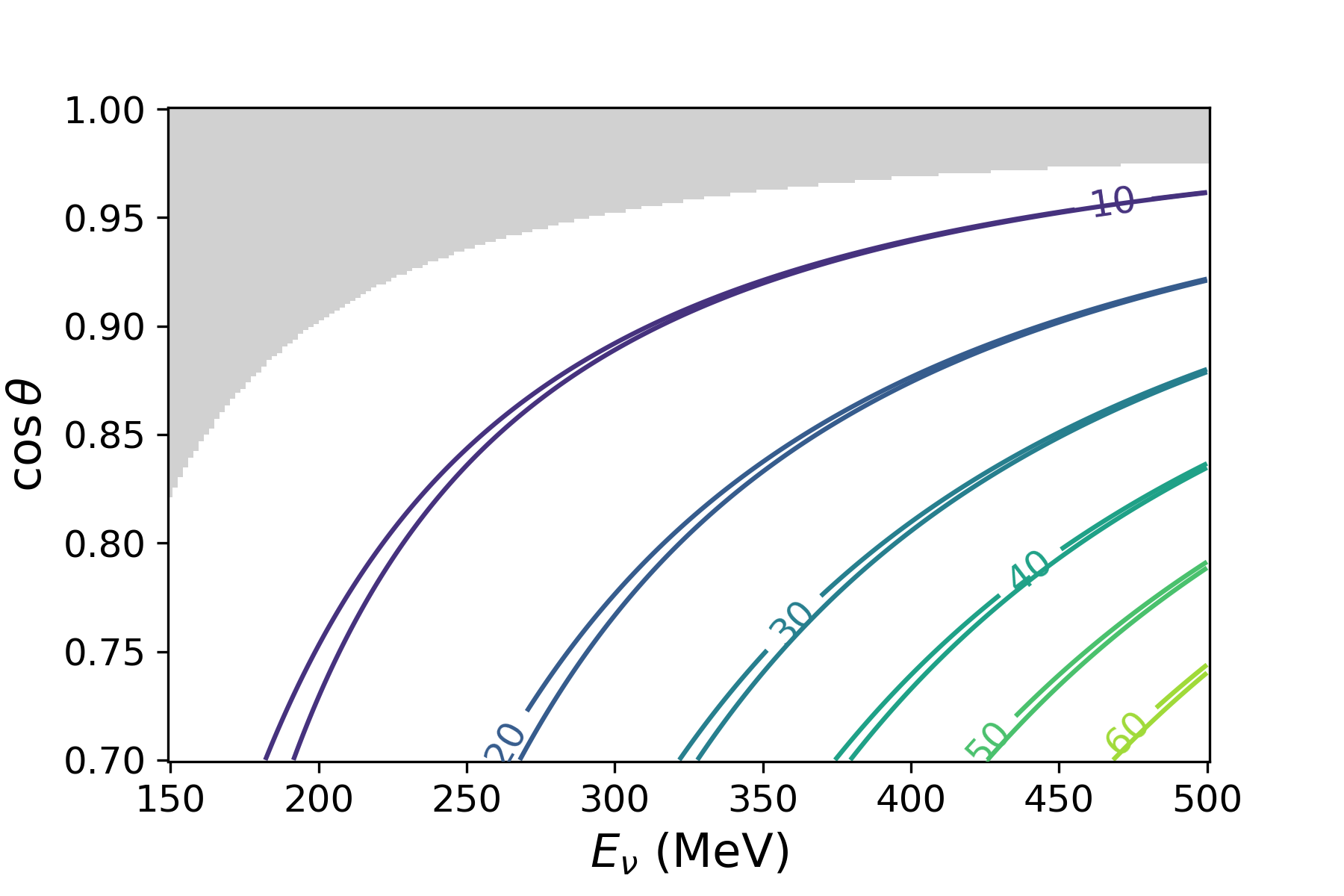}
\caption{Region, gray area, where $\omega_{QE}^{m_e}<\omega_{QE}^{m_\mu}$ in the $(\theta, E_{\nu_l})$ (left panel) and $(E_{\nu_l}, \cos\theta)$ (right panel) planes. Some constant values of $\omega_{QE}^{m_e}$ and $\omega_{QE}^{m_\mu}$ are represented by pair of continuous colored lines. The lines labeled by the $\omega_{QE}$ value (in MeV) are those corresponding to $\omega_{QE}^{m_e}$. }
\label{fig_plans_Enu_theta}
\end{center}
\end{figure}

Even if the figures are obtained by employing a peculiar approach, the Random Phase Approximation (RPA) one of Refs. \cite{Martini:2009uj,Martini:2011wp,Martini:2013sha}, general considerations can however be made. First of all, some well know features visually emerge:
\begin{itemize}
    \item The electron (anti) neutrino phase space is larger than the corresponding muon one, due to the different charged-lepton mass, which explains the larger total cross sections in the electron case. 
    \item The difference between the electron and muon (anti) neutrino cross sections decreases by increasing the neutrino energy. 
    \item The antineutrino cross sections decrease more rapidly increasing $q$, hence increasing the angle, than the neutrino ones. 
    \item The quasielastic response region clearly appears: all the cross sections are peaked at 
    %(or close to, due to RPA collective effects) 
    the quasielastic line \footnote{We remind that RPA collective effects may  shift the position of the QE peak, but the effect remains relatively weak. }
    \begin{equation}
    \label{eq_omega_qe}
       \omega_{QE}=\frac{q^2-\omega_{QE}^2}{2M_N}=\sqrt{q^2+M_N^2}-M_N 
    \end{equation}
     ($M_N$ being the nucleon mass) and spread around this curve due to Fermi motion.
\end{itemize}
The most important feature which emerges from the figures, and which represents one of the original results of the present work, concerns the inversion of the  $\nu_e$ ($\bar{\nu}_e$) and $\nu_\mu$ ($\bar{\nu}_\mu$) 
cross section hierachy and it is the following: 
\begin{itemize}
    \item At lower neutrino energies (for example $E_{\nu_l}=175$ MeV, as in Fig.\ref{fig_Enu175_mu_e_nu_anti}) the $\theta=0$ line largely crosses the quasielastic response region for the muon (anti) neutrino scattering, which is not the case of electron (anti) neutrino scattering, where the $\theta=0$ line is always outside the quasielastic response region. In other words, \textit{for neutrino and antineutrino  scattering the $\theta=0$ muon and electron lines explore in the $(q,\omega)$ plane two different regions, the muon one corresponding to larger quasielastic cross sections}. 
    %\MM{This is the simple, model-independent, reason of the $\nu_\mu$ over $\nu_e$ dominance, not formulated in these terms before. PEUT ETRE A DEPLACER.} 
    By increasing the neutrino energies the difference between the muon and electron $\theta=0$ lines decreases and the two curves explore more and more similar region in the $(q,\omega)$ plane, as it appears in Fig.\ref{fig_Enu575_mu_e_nu_anti}. 
\end{itemize}
The same argument allows us to see why at low neutrino energies the muon antineutrino cross sections are larger than electron ones also for backward scattering angles, as first observed in Ref.\cite{Dieminger:2023oin}: 
\begin{itemize}
    \item At
    low neutrino energies for antineutrino scattering the $\theta=\pi$ muon and electron lines explore in the $(q,\omega)$ plane two different regions, the muon one corresponding to larger quasielastic cross sections, as it appears in Fig.\ref{fig_Enu175_mu_e_nu_anti}. 
\end{itemize}

\begin{figure}
\begin{center}
  \includegraphics[width=16cm]{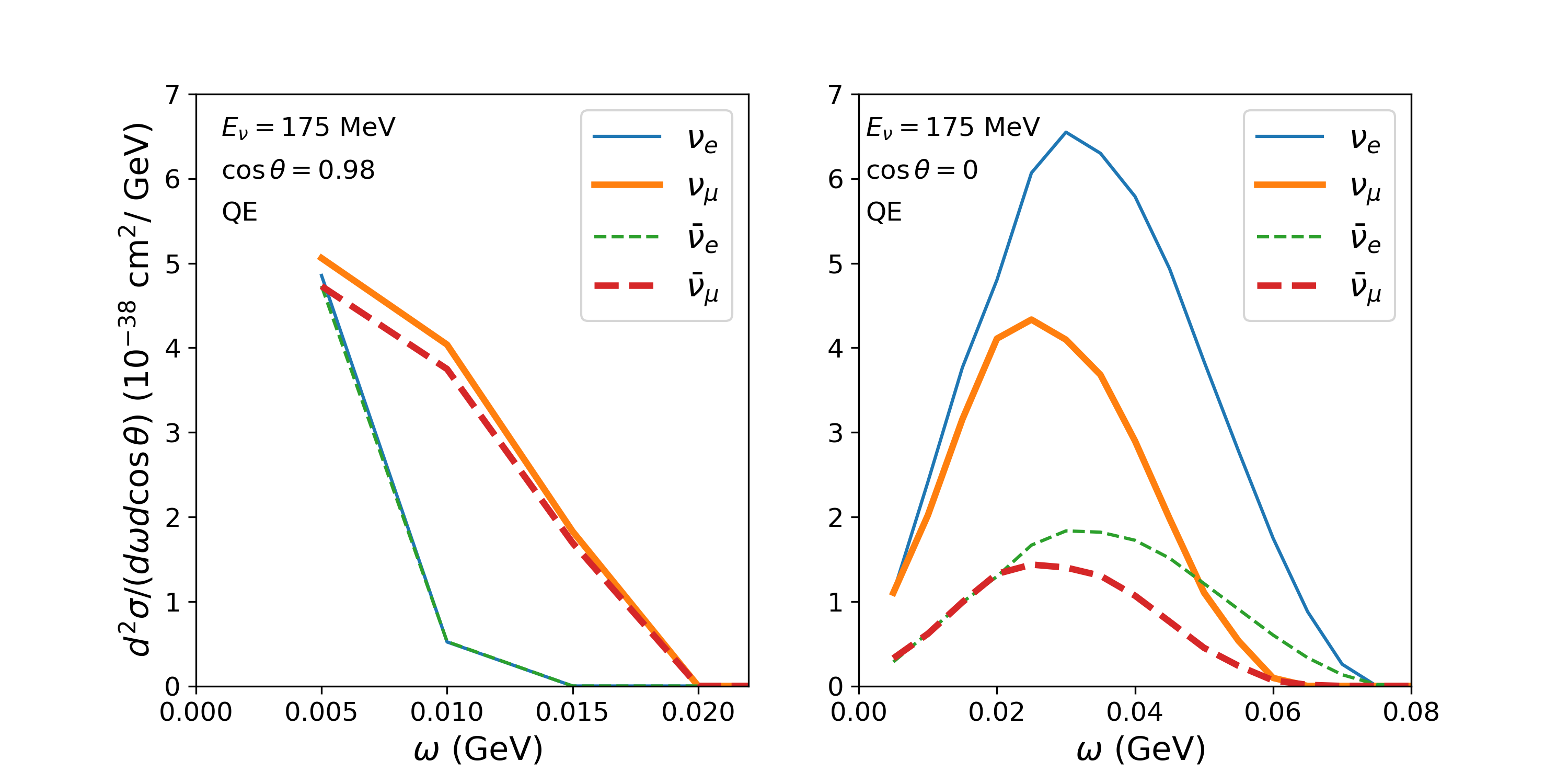}
\caption{The $\nu_e$, $\nu_\mu$, $\bar{\nu}_e$ and $\bar{\nu}_\mu$ charged current 
quasielastic double differential cross
sections on carbon $\frac{d^2 \sigma (E_{\nu_l})}{d\omega d\cos\theta}$  for the fixed value of the neutrino energy $E_{\nu_l}=175$ MeV and 
for two different values of the scattering angle.}
\label{fig_d2sqe_E175_cos098_cos0}
\end{center}
\end{figure}

To deepen our analysis let us consider now the energy position of the quasielastic peak in terms of the lepton variables $E_{\nu_l}$, $\cos\theta$ and $m_l$,
\begin{equation}
    \omega_{QE}^{m_l} \equiv \omega_{QE}(E_{\nu_l}, \cos\theta, m_l), 
\end{equation}
obtained by solving Eq.(\ref{eq_omega_qe}) once $q$ is expressed according to Eqs.(\ref{eq_q_relation}) and (\ref{eq_kl}). We omit to write its simple but long explicit expression, except in the case of zero charged-lepton mass, where it reduces to
\begin{equation}
\label{qe_line_ml0}
    \omega_{QE}^{m_l=0} = 
    \frac{E_{\nu_l}^2 (1-\cos\theta)}{M_N + E_{\nu_l} (1-\cos\theta)}.
\end{equation}
Depending on the values of $E_{\nu_l}$ and $\theta$, one of the two possibilities is realized: $\omega_{QE}^{m_e}<\omega_{QE}^{m_\mu}$ or $\omega_{QE}^{m_\mu}<\omega_{QE}^{m_e}$. This is illustrated in Fig.\ref{fig_plans_Enu_theta} which shows the $(\theta, E_{\nu_l})$ and the  $(E_{\nu_l}, \cos\theta)$ planes. 
%For the clarity of the figures we confine ourselves to relatively small scattering angles but we mention that, for the symmetry properties of $\omega_{QE}(E_{\nu_l}, \cos\theta, m_l)$, the same results are obtained by replacing $\theta$ with $\pi-\theta$, hence in the backward direction. 
In these planes the gray area delimits the region where $\omega_{QE}^{m_e}<\omega_{QE}^{m_\mu}$. One can observe that this situation occurs for small scattering angles and for very low values of $\omega_{QE}^{m_l}$, being $\omega_{QE}^{m_l} \lessapprox 5$ MeV. 
%As it is well known, and as it appears in Figs. \ref{fig_Enu175_mu_e_nu_anti} and \ref{fig_Enu575_mu_e_nu_anti}, at finite nuclear density there is a spread of the nuclear response around the quasielastic line $\omega_{QE}$ caused by the Fermi motion. 
For such very small values of $\omega_{QE}$, the region where $\omega \geq \omega_{QE}$ would mainly contribute to the response. As a consequence,  when  $\omega_{QE}^{m_e}<\omega_{QE}^{m_\mu}$ (or, in other words when the $\nu_\mu$ quasielatic peak is shifted at larger energies than the $\nu_e$ one) the contribution of the tail above the quasielastic peak will be  larger for $\nu_\mu$ than for $\nu_e$, hence  the $\nu_\mu$ quasielastic cross section will be larger than the $\nu_e$ one. This is illustrated in Fig. \ref{fig_d2sqe_E175_cos098_cos0} which shows the $\nu_e$, $\nu_\mu$, $\bar{\nu}_e$ and $\bar{\nu}_\mu$ charged current 
quasielastic double differential cross
sections on carbon $\frac{d^2 \sigma (E_{\nu_l})}{d\omega d\cos\theta}$  for $E_{\nu_l}=175$ MeV. The behaviour described above appears for $\cos\theta=0.98$, where $\omega_{QE}^{m_e}=0.65$ MeV is lower than $\omega_{QE}^{m_\mu}=1.24$ MeV \footnote{The introduction of an additional parameter to take into account the nucleon binding energy would shift the position of $\omega_{QE}$ towards larger energies but would not alter our conclusion.}. On the other hand, for $\cos\theta=0$,  $\omega_{QE}^{m_e}=27.49$ MeV is larger than $\omega_{QE}^{m_\mu}=22.48$ MeV and the electron (anti) neutrino cross sections are larger than the corresponding muon ones.

We have hence found the simple explanation on why the muon (anti) neutrino quasielastic cross sections can be larger than the electron ones: when $\omega_{QE}^{m_e}<\omega_{QE}^{m_\mu}$ the $\nu_\mu$ quasielastic cross sections ``falls after'' the $\nu_e$ ones. 
This simple explanation is strongly supported by the close correspondence between the left panel of our Fig.\ref{fig_plans_Enu_theta} and the left panel of Fig.4 of Ref.\cite{Nikolakopoulos:2019qcr}, 
both referring to the ($\theta,E_\nu$) plane: 
the region where $\omega_{QE}^{m_e}<\omega_{QE}^{m_\mu}$ (our Fig.\ref{fig_plans_Enu_theta}) practically coincides with the region where the calculations of the cross sections lead to $\frac{d\sigma_e/d\cos\theta}{d\sigma_\mu/d\cos\theta}<1$ (Fig.4 of Ref.\cite{Nikolakopoulos:2019qcr}).

\section{Other channels}

\begin{figure}
\begin{center}
  \includegraphics[width=16cm]{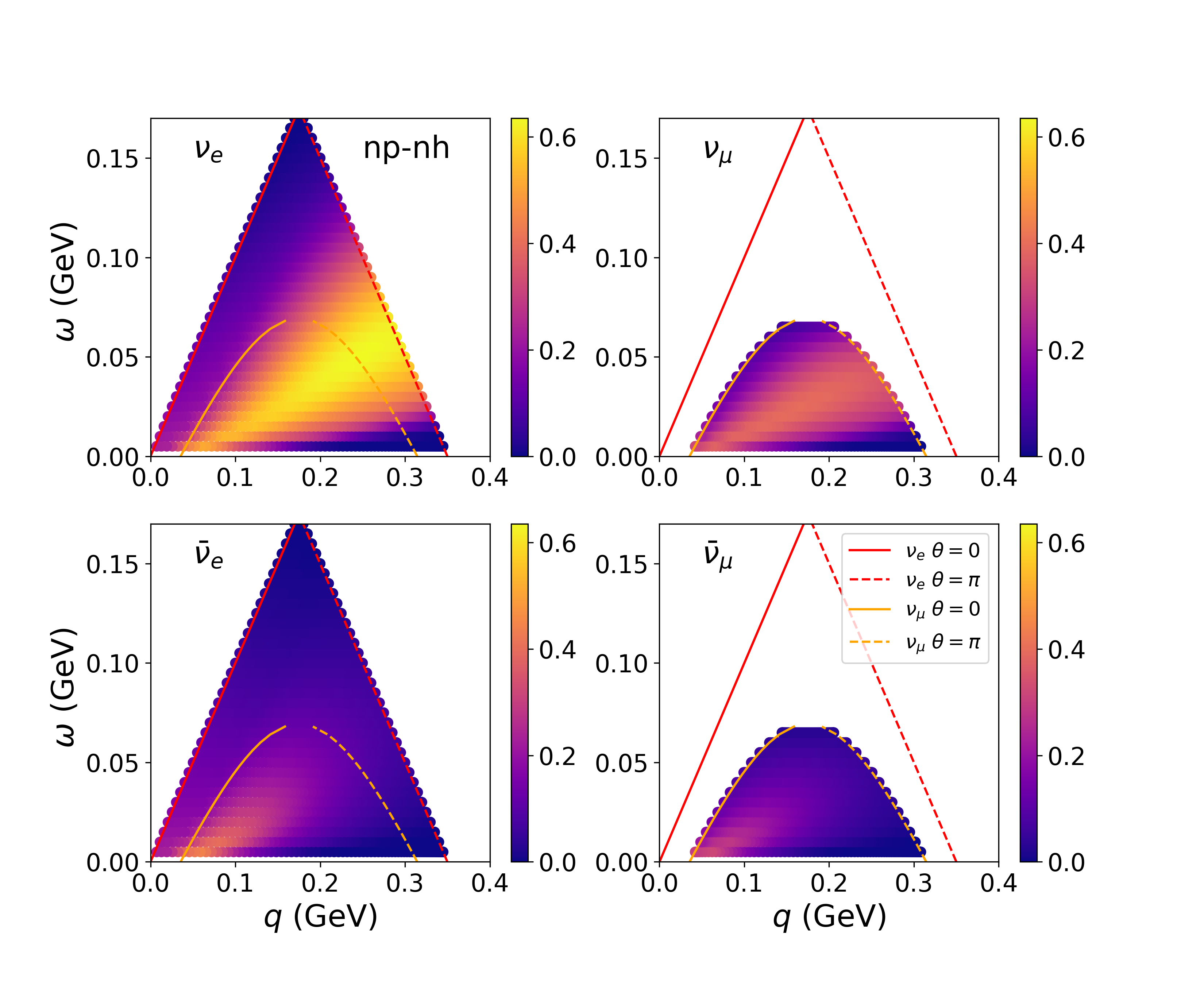}
\caption{The same as Fig.\ref{fig_Enu175_mu_e_nu_anti} at  $E_{\nu_l}=175$ MeV, but for np-nh excitations.}
\label{fig_npnh_Enu175_mu_e_nu_anti}
\end{center}
\end{figure}

\begin{figure}
\begin{center}
  \includegraphics[width=16cm]{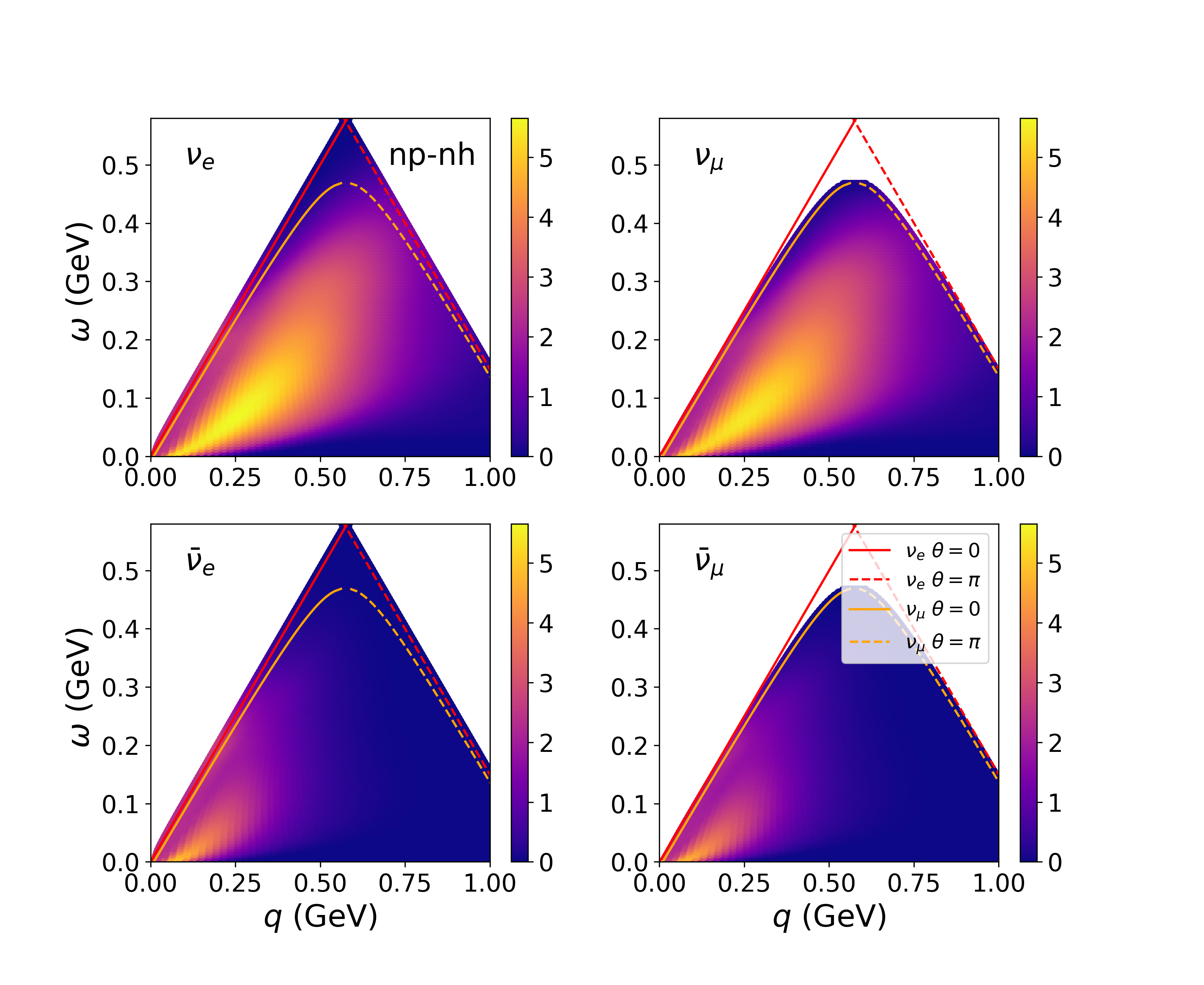}
\caption{The same as Fig.\ref{fig_Enu175_mu_e_nu_anti}, but for np-nh excitations and $E_{\nu_l}=575$ MeV.}
\label{fig_npnh_Enu575_mu_e_nu_anti}
\end{center}
\end{figure}

\begin{figure}
\begin{center}
  \includegraphics[width=16cm]{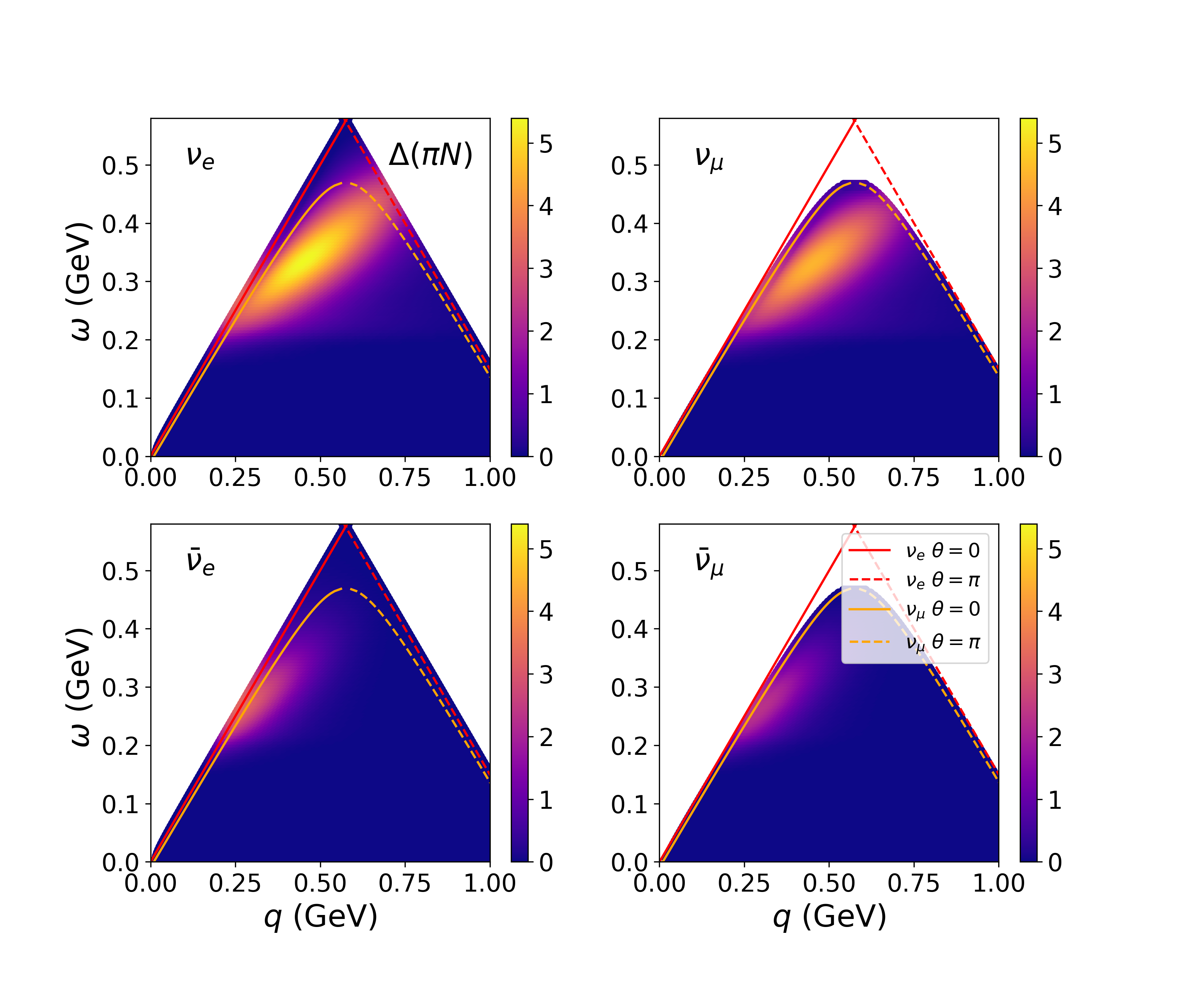}
\caption{The same as Fig.\ref{fig_Enu175_mu_e_nu_anti}, but for the 1 pion production via $\Delta$-resonance excitation and $E_{\nu_l}=575$ MeV.}
\label{fig_delta_Enu575_mu_e_nu_anti}
\end{center}
\end{figure}

\begin{figure}
\begin{center}
  \includegraphics[width=16cm]{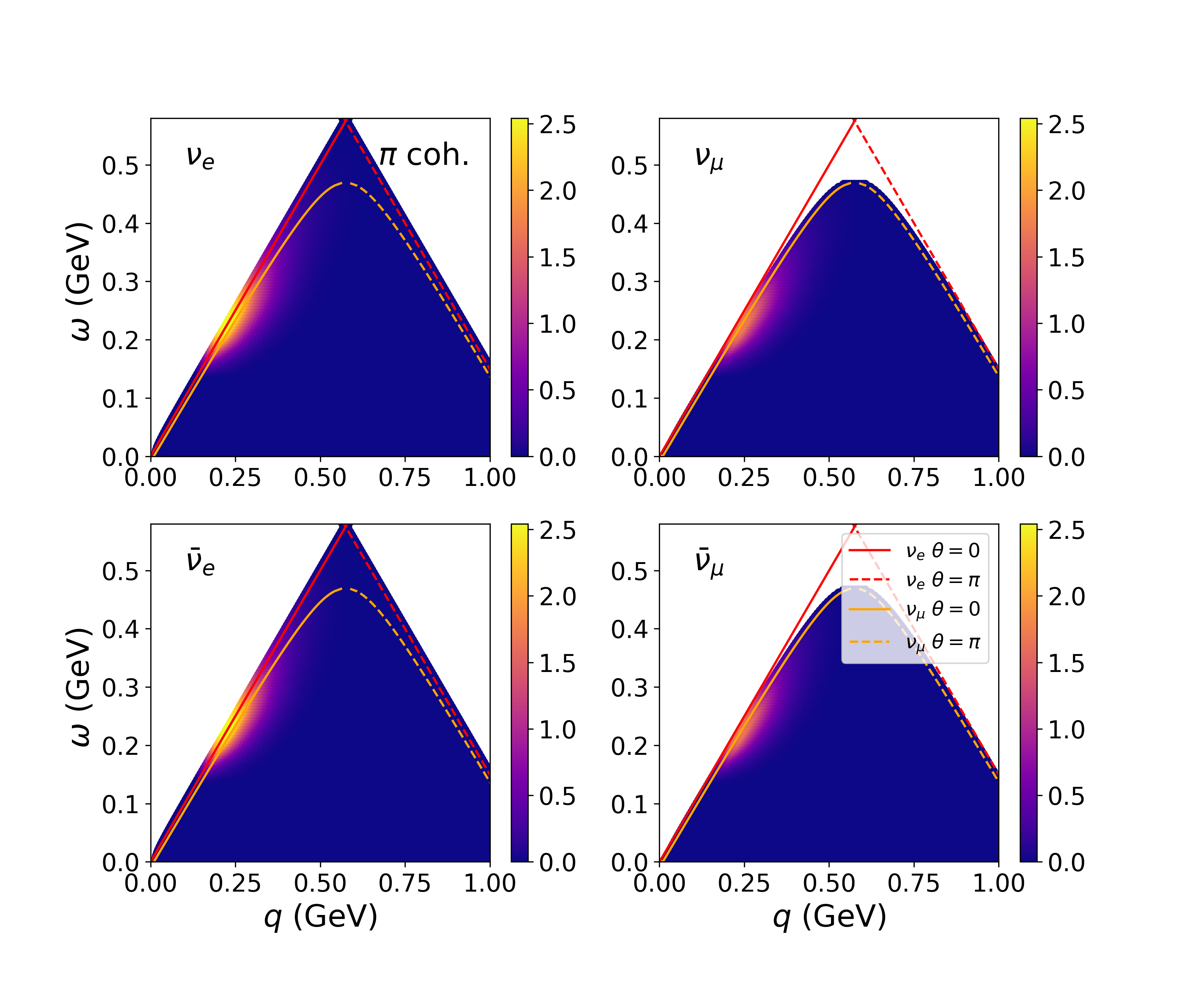}
\caption{The same as Fig.\ref{fig_Enu175_mu_e_nu_anti}, but for the coherent 1 pion production and $E_{\nu_l}=575$ MeV.}
\label{fig_pi_coh_Enu575_mu_e_nu_anti}
\end{center}
\end{figure}

Let us move now to the other channels described by our model, the multinucleon emission and the resonant and coherent one pion production, and let us start by showing, as for the quasielastic, the projection of the corresponding double differential cross sections on the $(q,\omega)$ plane. The case of multinucleon emission, including 2p-2h and 3p-3h excitations, hence called np-nh, is shown in Fig.\ref{fig_npnh_Enu175_mu_e_nu_anti} for $E_{\nu_l}=175$ MeV and in Fig.\ref{fig_npnh_Enu575_mu_e_nu_anti} for $E_{\nu_l}=575$ MeV. For this last value of the neutrino energy, the resonant 1 pion production result is shown in Fig.\ref{fig_delta_Enu575_mu_e_nu_anti} and the coherent pion one in Fig.\ref{fig_pi_coh_Enu575_mu_e_nu_anti}. 

From Figs.\ref{fig_npnh_Enu175_mu_e_nu_anti} and \ref{fig_npnh_Enu575_mu_e_nu_anti} one can observe that in the case of np-nh excitations the cross section, which reflects the nuclear response region, is not
restricted to the Fermi motion band around the quasielastic line
(as in Figs. \ref{fig_Enu175_mu_e_nu_anti} and \ref{fig_Enu575_mu_e_nu_anti}) but it covers the major part of the $(q,\omega)$ plane. As a consequence the intersection between the $\theta=0$ line and the nuclear response, does not seem to clearly indicate the dominance of muon (anti) neutrino cross section over the electron ones at low neutrino energy, as for quasielastic. Some cases where this phenomenon locally survives are discussed later.

Concerning the incoherent 1 pion production cross sections projected in the $(q,\omega)$ plane in Fig.\ref{fig_delta_Enu575_mu_e_nu_anti}, one can recognize the $\Delta$-resonance response region peaked around the $\Delta$-line 
\begin{equation}
\label{delta_disp_rel}
\omega_{\Delta}=\sqrt{q^2+M_{\Delta}^2}-M_N. 
\end{equation}
 By ignoring the charged-lepton mass, Eq.\eqref{delta_disp_rel} can be written as: 
\begin{equation}
\label{delta_line}
\omega_{\Delta}^{m_l=0}=\frac{M_N{ \Delta  M +E_{\nu_l}}^2 (1-\cos\theta)}{M_N + E_{\nu_l} (1-\cos\theta)}, 
\end{equation}
with $ \Delta  M =(M^2_{\Delta}-M^2_N)/2M_N= 338$ MeV. Equation (\ref{delta_line}) is 
the corresponding of Eq.(\ref{qe_line_ml0}), related to quasielastic, in the case of $\Delta$. The spread around the $\Delta$ line is due to the in-medium $\Delta$ width and to the Fermi motion. 

Turning to the projection in the $(q,\omega)$ plane of the coherent 1 pion production cross sections, shown in  Fig.\ref{fig_pi_coh_Enu575_mu_e_nu_anti}, one can observe the accumulation of strength around the $\theta=0$ line, which coincide with the $\omega=q$ line when $m_l=0$, reflecting the free pion dispersion relation
\begin{equation}
   \omega_\pi=\sqrt{q^2+m_\pi^2}. 
\end{equation}

\begin{figure}
\begin{center}
  \includegraphics[width=16cm]{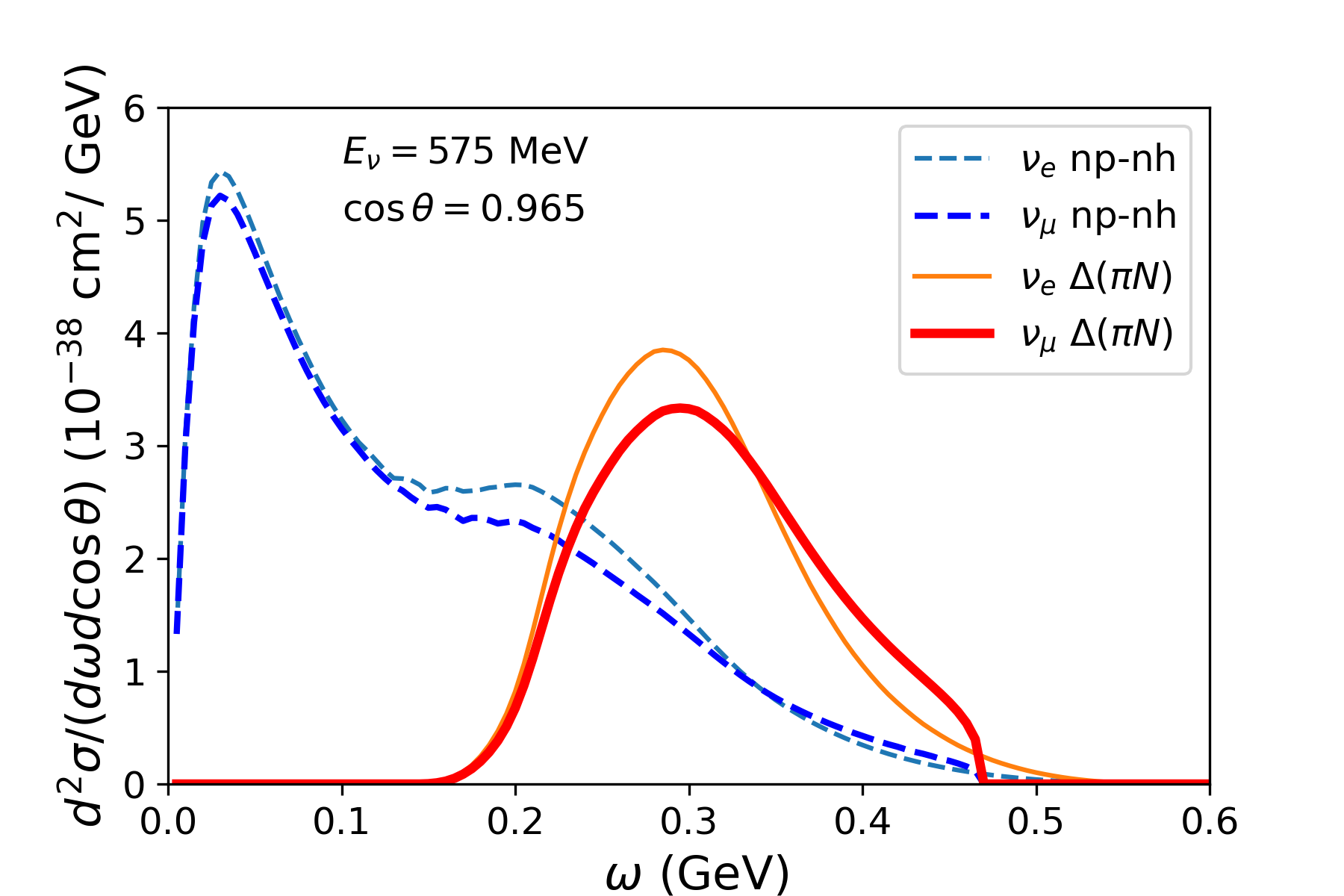}
\caption{The $\nu_e$ and $\nu_\mu$ double differential cross
sections on carbon $\frac{d^2 \sigma (E_{\nu_l})}{d\omega d\cos\theta}$ 
for np-nh exciations and for the resonant one pion production at $E_{\nu_l}=575$ MeV and  $\cos\theta=0.965$.}
\label{fig_d2s_npnh_D_E575_cos0965}
\end{center}
\end{figure}

\begin{figure}
\begin{center}
  \includegraphics[width=16cm]{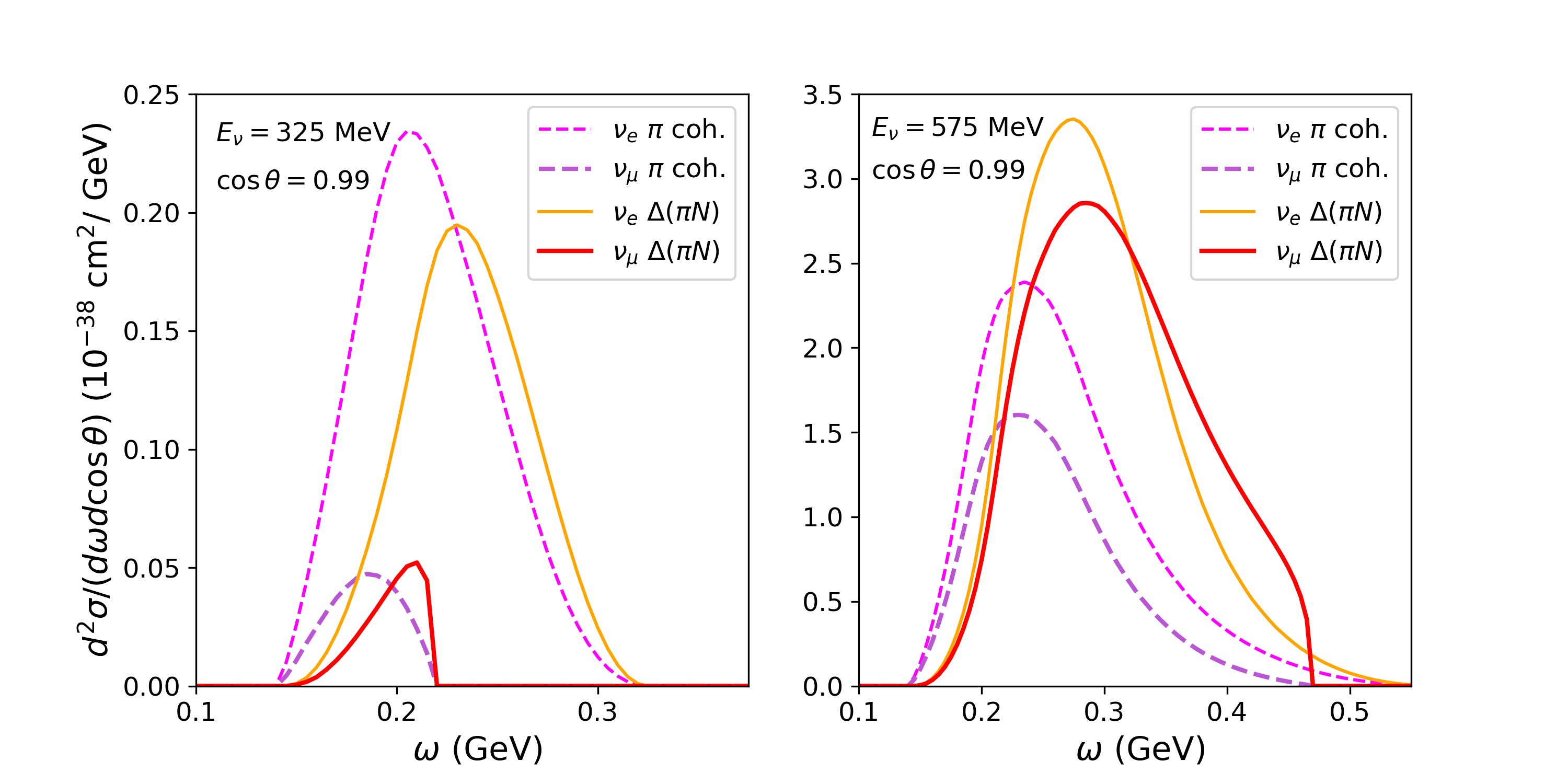}
\caption{The $\nu_e$ and $\nu_\mu$ double differential cross
sections on carbon $\frac{d^2 \sigma (E_{\nu_l})}{d\omega d\cos\theta}$ 
for coherent and resonant one pion production at $E_{\nu_l}=325$ MeV and $E_{\nu_l}=575$ MeV and  $\cos\theta=0.99$.}
\label{fig_d2s_coh_D_E325_575_cos099}
\end{center}
\end{figure}

From Figs.\ref{fig_delta_Enu575_mu_e_nu_anti} and \ref{fig_pi_coh_Enu575_mu_e_nu_anti} the resonant and coherent one pion production (anti) neutrino double differential cross sections seem to be globally larger in the electron than in the muon case. It is however interesting to consider these cross sections for some fixed kinematics as a function of the transferred energy to investigate the possible emergence of some region where this hierarchy is opposite as well the appearance of other effects. For this purpose we plot in Fig.\ref{fig_d2s_npnh_D_E575_cos0965} this cross section at $E_{\nu_l}=575$ MeV and  $\cos\theta=0.965$ for the resonant pion production channel. At this kinematics the position of the $\Delta$ peak is at larger $\omega$ for $\nu_\mu$ than for $\nu_e$, hence an effect analog to the one discussed above for quasielastic could in principle appears: the spread of the $\Delta$ width could shift the $\nu_\mu$ cross section at larger $\omega$ leading to a larger cross section for $\nu_\mu$ than for $\nu_e$ in the tail. Indeed this is what locally happens. However the effects is less evident than for the quasielastic. Furthermore it is combined to threshold effects related to the finite muon mass, leading to peculiar shape of the $\nu_\mu$ cross section near the maximum value of the allowed excitation energy. The same threshold effect, as well as the dominance of the $\nu_\mu$ over $\nu_e$ is also visible, even if weakly, in the tail of multinucleon cross section (plotted in Fig.\ref{fig_d2s_npnh_D_E575_cos0965} as well), the np-nh being due to the non pionic decay of the $\Delta$ in this tail.

Figure \ref{fig_d2s_coh_D_E325_575_cos099} shows the coherent and resonant one pion production double differential cross sections at $\cos\theta=0.99$ for two values of neutrino energy, $E_{\nu_l}=325$ MeV and $E_{\nu_l}=575$ MeV. Beyond to retrieve for the resonant cross section at $E_{\nu_l}=575$ MeV the behavior already discussed for Fig.\ref{fig_d2s_npnh_D_E575_cos0965}, one can observe that the coherent and the resonant cross sections are peaked at different energies, reflecting the differences between $\omega_\pi$ and $\omega_\Delta$. Furthermore at $E_{\nu_l}=325$ MeV the maximum value of the cross section is larger for the coherent than for the resonant channel in the case of $\nu_e$ scattering.

\section{Summary and Conclusion}
In summary we have explained why the $\nu_\mu$ quasielastic differential cross sections can be larger than the corresponding $\nu_e$ ones 
by analyzing the $(q,\omega)$ and the $(\theta,E_\nu)$ phase spaces. 
We have found a simple criterium to determine when muon (anti) neutrino quasielastic differential cross sections are larger than the corresponding electron ones, based on the position of the quasielastic peak at very low transferred energy.  This criterium, which does not need the explicit calculation of the cross sections, could be useful for experimental analyses allowing simple cuts to exclude regions where the modeling cross section is expected to be not so robust.  

Also in the one pion production and in the  multinucleon emission channels, for peculiar kinematical conditions, we have found in the tails of the double differential cross sections as a function of the transferred energy a dominance of $\nu_\mu$ over the $\nu_e$ results. The shape of these tails is also affected by threshold effects. 

As a perspective, even if the correspondence between $\nu_e$ and $\nu_\mu$ cross section is not trivial,  one could use the representation of the cross sections in terms of the $q$ and $\omega$ variables, as done in this work, to investigate if it is possible to find patterns allowing to constrain unmeasured electron (anti)neutrino cross sections starting from the measured muon ones. 

\section*{Acknowledgement}
We thank T. Dieminger, S. Dolan, C. Giganti, L. Russo, Y. Maidannyk, D. Sgalaberna and U. Virginet for useful discussions. 
\bibliography{biblio}
\end{document}